\documentclass{acm_proc_article-sp}

\begin{document}
\sloppy
\title{Job-related discourse on social media}

\numberofauthors{6} %
\author{
\alignauthor
Tong Liu\\
    \affaddr{Department of Computer Science}\\
    \affaddr{Rochester Institute of Technology}\\
    \affaddr{Rochester, NY}\\
    \email{tl8313@rit.edu}
\alignauthor
Christopher M. Homan\\
    \affaddr{Department of Computer Science}\\
    \affaddr{Rochester Institute of Technology}\\
    \affaddr{Rochester, NY}\\
    \email{cmh@cs.rit.edu}
\alignauthor 
Cecilia Ovesdotter Alm\\
    \affaddr{Department of English}\\
    \affaddr{Rochester Institute of Technology}\\
    \affaddr{Rochester, NY}\\
    \email{coagla@rit.edu}
\and  %
\alignauthor 
Ann Marie White\\
    \affaddr{Department of Psychiatry}\\
    \affaddr{University of Rochester Medical Center}\\
    \affaddr{Rochester, NY}
\alignauthor 
Megan C. Lytle-Flint\\
    \affaddr{Department of Psychiatry}\\
    \affaddr{University of Rochester Medical Center}\\
    \affaddr{Rochester, NY}
\alignauthor 
Henry A. Kautz\\
    \affaddr{Goergen Institute for Data Science}\\
    \affaddr{University of Rochester}\\
    \affaddr{Rochester, NY}
}

\maketitle

\begin{abstract}
Working adults spend nearly one third of their daily time at their jobs. In this paper, we study job-related social media discourse from a community of users. We use both crowdsourcing and local expertise to train a classifier to detect job-related messages on Twitter. Additionally, we analyze the linguistic differences in a job-related corpus of tweets  between individual users vs. commercial accounts. The volumes of job-related tweets from individual users indicate that people use Twitter with distinct monthly, daily, and hourly patterns. We further show that the moods associated with jobs, positive and negative, have unique diurnal rhythms.
\end{abstract}




\section{Introduction}
In this paper, we build a robust language-based classifier that can automatically and accurately identify Twitter messages about job-related topics. The classifier is trained by a supervised learning pipeline that uses humans-in-the-loop to boost model performance. We discover and analyze temporal patterns in the volume of job-related tweets and investigate how positive and negative affect in job-related tweets vary over time. %

The obvious importance of this research is that working-age Americans spend on average 8.7 hours per day in job-related activities \cite{timeuse}, which is more time than any other single activity. So any attempt to understand a working individual's experiences, state of mind, or motivations must take into account their life at work. Social media has become a rich source of \emph{in situ} data for research on social issues and behaviors, yet to our knowledge, little of this work has focused on how individuals talk there about their jobs.

Conversely, a better understanding of how people discuss work informally through social media can potentially shed light on behavioral problems that impact work. 70\% of US workers are disengaged at work \cite{gallup}. This hurts companies and costs  between 450 and 550 billion dollars each year in lost productivity. Disengaged workers are 87\% more likely to leave their jobs than their more satisfied counterparts \cite{gallup}. 
Job dissatisfaction poses serious health risks and even leads to suicide \cite{worksuicide}. The deaths by suicide among working age people (25-64 years old) costs more than \$44 billion annually \cite{suicidefigure}. 

The need for machine learning, rather than simple heuristics, to discover work-related social media posts is due to the inherent ambiguity of language related to jobs. For instance, a tweet like ``\emph{@SOMEONE @SOMEONE shit manager shit players shit everything}'' contains the work-related word ``manager,'' yet the presence of ``player'' ultimately suggests this tweet is about a sport team. The tweet ``\emph{@SOMEONE anytime for you boss lol}'' might seem job-related, but ``boss'' here could also simply mean ``friend'' in an informal and familiar register.

Aggregated job-related information from Twitter can be valuable to a range of stakeholders. For instance, 
public health specialists, psychologists and psychiatrists could use such first-hand reportage of work experiences to monitor job-related stress at a community level. Employers might use it to improve how they manage their business, cost, and entrepreneurial energy. It could also help employees to maintain better online reputations for potential job recruiters. We also recognize that there are ethical considerations involved for analyzing job-related distress of individuals (e.g., supervisors monitoring particular employees' job satisfaction).

\section{Background and Related Work}
Social media accounts for about 20\% of the time spent online \cite{socialtop}. Online communication has a faceless nature that can embolden people to reveal their cognitive state in a natural, unselfconscious manner \cite{iKeepSafe}. Mobile phone platforms help social media to capture personal behavior \emph{in situ}, whenever and wherever possible \cite{de2013predicting,sadilek2013modeling}. These signals are often temporal, and can reveal how phenomena change over time. Thus, aspects about individuals or groups, such as preferences and perspectives, affective states and experiences, communicative patterns, and socialization behaviors can, to some degree, be analyzed and computationally modeled continuously and unobtrusively \cite{de2013predicting}.

Previous research shows that social media can predict political inclination \cite{tumasjan2010predicting}, or performance of stock market \cite{bollen2011twitter}, which reflect the social and economic situations. The spread of infectious diseases, like flu, can also be predicted through online social media \cite{sadilek2012modeling}. Syndromic surveillance system for multiple ailments also can be established with social media data \cite{paul2011you}. Smoking and drinking, depression, domestic abuse, and other behavioral and public wellness problems can also be analyzed \cite{tamersoy2015characterizing,de2013predicting,schrading2015whyistayed}.

In contrast to such prior studies, we focus on a broad discourse and narrative theme that touches most adults. Measures of volume, content, affect of job-related discourse on social media may help understand the behavioral patterns of working people, predict labor market changes, monitor and control satisfaction/dissatisfaction with respect to their workplaces or colleagues, and help people strive for positive change \cite{de2013understanding}. The language differences exposed in social media have been observed and analyzed in relation to location \cite{cheng2010you}, gender, age, regional origin, and political orientation \cite{rao2010classifying}. 
With the help of social media, researchers have identified individual-level diurnal and seasonal mood rhythms in cross-culture comparisons \cite{golder2011diurnal}. We are interested to know the affective rhythms hidden in job-related discourse on social media specifically.

Twitter has drawn much attention from researchers in various disciplines in large part because of the volume and granularity of publicly available social data. This micro-blogging website, which was launched in 2006, has attracted more than 500 million registered users by 2012, with 340 million tweets posted every day. Twitter supports directional connections (followers and followees) in its social network, and allows for geographic information about where a tweet was posted if a user enables location services. In our context, this nature of the data allows us to study job-related topics in multidimensional ways.

LIWC\footnote{Linguistic Inquiry and Word Count, http://www.liwc.net/}, has proved useful for extracting the psychological dimensions of language \cite{tausczik2010psychological} and address many challenging problems, such as classifying depression and paranoia sufferers \cite{oxman1982language}, monitoring emotion expression under stress in instant messaging \cite{pirzadeh2012emotion}, characterizing sentiment in tweets \cite{pak2010twitter,nasukawa2003sentiment}, revealing cues about neurotic tendencies and psychiatric disorders \cite{rude2004language}, and estimating the risk of suicide from  unstructured clinical records \cite{poulin2014predicting}. 

\section{Research Questions}

Here, we pursue the following questions:
\begin{description}
\item[RQ 1:]
Can we identify linguistic characteristics that differentiate groups of users who talked about job-related topics? 

\item[RQ 2:] What are job-related topics usually about?

\item[RQ 3-1:]
How do posts of job-related messages change over the course of a year?

\item[RQ 3-2:]
From Monday to Sunday in each week, what patterns emerge? For instance, on which day do people talk the most vs. the least about job-related topics? 

\item[RQ 3-3:]
When is the most active period in a day when people tweet about jobs?

\item[RQ 4:]
How do tweeters' affective tone change over time in job-related tweets? Are there observable variations by seasonal or diurnal factors?
\end{description}

\section{Data and Methods}

\begin{figure*}[ht]
\includegraphics[width=\linewidth]{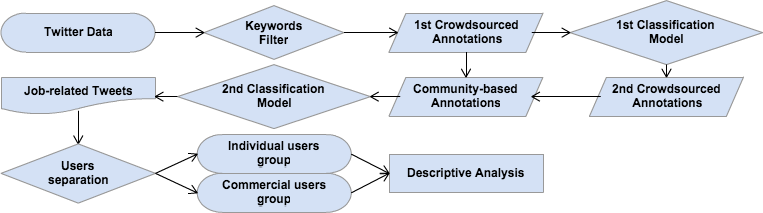}
\caption{Overview of our detection and analysis framework}\label{workflow}
\centering
\end{figure*}

In this section, we first review our method of building an iterative humans-in-the-loop supervised learning framework to automatically detect the job-related messages from Twitter \cite{tong2015detecting}. Then we leverage the pipeline and labeled tweets and developed a series of descriptive analysis. Figure \ref{workflow} summarizes our workflow.

Using the DataSift\footnote{http://datasift.com/} Firehose, we collected tweets from public accounts with geographical coordinates located in a 15-counties region surrounding a mid-sized US city from July 2013 to June 2014. This data set contains over 7 million geo-tagged tweets (approximately 90\% written in English) from around 85,000 unique Twitter accounts.
We fix our data to this particular locality because it is geographically diverse, covering both urban and rural areas and providing mixed and balanced demographics. Also, due to the nature of the subject matter, it is helpful to use knowledge about the local job scene in the modeling and analysis. 

Then, to preprocess the tweets, we remove punctuation and special characters, and heuristically map informal terms to standard ones using the Internet Slang Dictionary\footnote{http://www.noslang.com/dictionary}. We also remove special characters, like emoticons, before conducting a crowdsourcing study.

\subsection{Data filtering}
Words such as ``job'' have multiple meanings. In order to identify likely job-related tweets while excluding others (such as those discussing homework or other school-related activities) we filtered the tweets using the inclusion and exclusion terms shown in Table \ref{table:1}. This yielded over 40,000 tweets  %
having at least five tokens each. These tweets were labeled \emph{Job-Likely}.

\begin{table}[h]
\centering 
\begin{tabular}{|c|c|}
\hline
\multirow{2}{*}{\textbf{Include}} & job, jobless, manager, boss \\ \cline{2-2} 
 & my/your/his/her/their/at work \\ \hline
\multirow{2}{*}{\textbf{Exclude}} & school, class, homework, student, course \\ \cline{2-2} 
 & good/nice/great job \\ \hline
\end{tabular}
\caption{Filters used to extract the \emph{Job-Likely} set.}
\label{table:1}
\end{table}

\subsection{Crowdsourced annotation}

We randomly chose around 2,000 \emph{Job-Likely} tweets and split them equally into 50 subsets of 40 tweets each. To measure both inter- and intra-annotator agreement, we additionally randomly duplicated five tweets in each subset. We then constructed Amazon Mechanical Turk (AMT)\footnote{https://www.mturk.com/mturk/welcome} Human Intelligence Tasks (HITs) to collect reference annotations. For each tweet, we asked workers, ``\emph{Is this tweet about employment or job?}''. The answer ``Y'' means ``job-related'' and ``N'' means ``not job-related''.

We assigned five crowdworkers to each HIT -- this is an empirical scale for crowdsourced linguistic annotation tasks suggested by \cite{callison2009fast,evanini2010using}. Crowdworkers were required to live in the United States and have an approval rating of 90\% or better. They were paid \$1.00 per HIT. Workers were allowed to work on as many distinct HITs as they liked, and bonuses were given to those who completed multiple HITs. To evaluate annotation quality, we examined whether each worker provided identical answers to the five duplicate tweets. Among the annotators of each HIT, we calculated Fleiss' kappa \cite{fleiss1971measuring} and Krippendorff's alpha \cite{klaus1980content} measures, using the tool\footnote{Inter-Rater Agreement with multiple raters and variables, https://mlnl.net/jg/software/ira/} to assess inter-annotator reliability. Our conjecture, borrowed from \cite{snow2008cheap}, is that labeled tweets with high inter-annotator agreement among crowdworkers can be used to build a robust model. The above measures also help us decide whether to reward an annotator in full or partially. 

Before publishing the HITs, we also consulted with Turker Nation\footnote{http://www.turkernation.com} to ensure that the workers were treated and compensated fairly for their tasks.

\subsection{Classification model}

The aforementioned labeling task yielded 1,297 tweets where all five annotators agreed on the labels. 1,027 of these were labeled ``job-related'' (and the rest ``not job-related''). To construct a balanced training set, we added another 757 tweets chosen randomly from tweets outside the \emph{Job-Likely} set. 

After converting text to lower case, text features were extracted as unigrams, bigrams, and trigrams. For example, the tweet ``\emph{I really hate my job}'' is represented as \{i, really, hate, my, job, i really, really hate, hate my, my job, i really hate, really hate my, hate my job\}. SVM$^{light}$\footnote{http://svmlight.joachims.org} was used to train the classification model, which was then used to classify the rest of the dataset. Excluding tweets with less than five tokens, the model labeled 
a total of 535,646 tweets as job-related and 4,465,616 tweets as not.

\subsection{Second crowdsourced annotation}

To generate better training data and evaluate the effectiveness of the aforementioned model, a second round of labeling was conducted. This assigned to each tweet in the dataset a confidence score, defined as its distance to separating hyperplanes determined by the support vectors. After separating positive- and negative-labeled (job-related vs. not) tweets, each group was sorted in descending order of their confidence scores.

We used about 4,000 of these sorted tweets in the second round of AMT HITs. Part of these tweets were randomly chosen from the subset of the positive class, with the 80th percentile of confidence scores as a cutoff point for inclusion (labeled as \emph{Type-1} in Table \ref{round2_model_human}). The rest were obtained from those tweets in either class having confidence scores close to zero (\emph{Type-2}). This latter set represents those tweets that are ambiguous and ``difficult'' for the classifier to label. Hence, we consider both the clearer core and at the gray zone periphery of this meaning phenomenon, which adds an interesting challenge to the classification process. %
Table \ref{round2_model_human} records how these two types of tweets were annotated.

\begin{table}[ht]
\centering
\begin{tabular}{|c|c|c|c|c|c|c|}
\hline
\multirow{3}{*}{\textbf{Round 2}} & \multicolumn{6}{c|}{\textbf{\begin{tabular}[c]{@{}c@{}}Number of agreements\\  among five annotators\end{tabular}}} \\ \cline{2-7} 
 & \multicolumn{3}{c|}{\textbf{job-related}} & \multicolumn{3}{c|}{\textbf{not job-related}} \\ \cline{2-7} 
 & \textbf{3} & \textbf{4} & \textbf{5} & \textbf{3} & \textbf{4} & \textbf{5} \\ \hline
\textbf{Type-1} & 129 & 280 & 713 & 50 & 149 & 1079 \\ \hline
\textbf{Type-2} & 11 & 7 & 8 & 16 & 67 & 1489 \\ \hline
\end{tabular}
\caption{Summary of the two types of tweets in the second crowdsourced annotation and the corresponding annotations}
\label{round2_model_human}
\end{table}

Table \ref{table:3} summarizes the results from both annotation rounds. 

\begin{table}[h]
\centering
\begin{tabular}{|c|c|c|c|c|c|c|}
\hline
\multirow{3}{*}{\textbf{Round 1+2}} & \multicolumn{6}{c|}{\textbf{\begin{tabular}[c]{@{}c@{}}Number of agreements\\ among five annotators\end{tabular}}} \\ \cline{2-7} 
 & \multicolumn{3}{c|}{\textbf{job-related}} & \multicolumn{3}{c|}{\textbf{not job-related}} \\ \cline{2-7} 
 & \textbf{3} & \textbf{4} & \textbf{5} & \textbf{3} & \textbf{4} & \textbf{5} \\ \hline
\textbf{Round 1} & 104 & 389 & 1027 & 78 & 116 & 270 \\ \hline
\textbf{Round 2} & 140 & 287 & 721 & 66 & 216 & 2568 \\ \hline
\end{tabular}
\caption{Summary of both annotation rounds}
\label{table:3}
\end{table}

Table \ref{samples} displays all the inter-annotator agreement combinations among five annotators and sample tweet in each case (selected from both annotation rounds).

\begin{table}[h]
\centering
\begin{tabular}{|c|c|}
\hline
\textbf{\begin{tabular}[c]{@{}c@{}}Crowdsourced\\ Annotations\\ Y/N\end{tabular}} & \textbf{Sample Tweet} \\ \hline
\textbf{Y, Y, Y, Y, Y} & \begin{tabular}[c]{@{}c@{}}Really bored....., no entertainment\\ at work today\end{tabular} \\ \hline
\textbf{Y, Y, Y, Y, N} & \begin{tabular}[c]{@{}c@{}}two more days of work then\\ I finally get a day off.\end{tabular} \\ \hline
\textbf{Y, Y, Y, N, N} & \begin{tabular}[c]{@{}c@{}}Leaving work at 430 and\\ driving in this snow is going\\ to be the death of me\end{tabular} \\ \hline
\textbf{Y, Y, N, N, N} & \begin{tabular}[c]{@{}c@{}}Being a mommy is the hardest\\ but most rewarding job\\ a women can have\\ \#babyBliss \#babybliss\end{tabular} \\ \hline
\textbf{Y, N, N, N, N} & \begin{tabular}[c]{@{}c@{}}These refs need to\\ DO THEIR FUCKING JOBS\end{tabular} \\ \hline
\textbf{N, N, N, N, N} & \begin{tabular}[c]{@{}c@{}}One of the best Friday nights\\ I've had in a while\end{tabular} \\ \hline
\end{tabular}
\caption{Inter-annotator agreement combinations and sample tweets}
\label{samples}
\end{table}

\subsection{Community-based annotation}

The job-related salience of those tweets in which the majority --- but not all --- of the annotators agreed (i.e., 3 or 4 out of 5) is less clear than of those with unanimous agreement, but such less-clear tweets are still potentially useful. Integrating four subsets of tweets from both rounds of crowdsourced annotations --- (a) tweets with only 3 crowdworkers answered ``Y'' (referred in Table \ref{table:5} as \emph{job-3}); (b) tweets with only 3 crowdworkers answered ``N'' (\emph{not-job-3}); (c) tweets with only 4 crowdworkers answered ``Y'' (\emph{job-4}); and (d) tweets with only 4 crowdworkers answered ``N'' (\emph{not-job-4}) --- we asked two co-authors from the local community to also review them and provide a gold-standard label. Table \ref{table:5} summarizes results from this phase.

\begin{table}[h]
\centering
\begin{tabular}{|c|c|c|}
\hline
\multirow{2}{*}{\textbf{Round 1+2}} & \multicolumn{2}{c|}{\textbf{\begin{tabular}[c]{@{}c@{}}Annotations collected \\ from the local community\end{tabular}}} \\ \cline{2-3} 
 & \textbf{job-related} & \textbf{not job-related} \\ \hline
\textbf{job-3} & 197 & 21 \\ \hline
\textbf{not-job-3} & 62 & 63 \\ \hline
\textbf{job-4} & 651 & 11 \\ \hline
\textbf{not-job-4} & 12 & 317 \\ \hline
\end{tabular}
\caption{Summary of community-based reviewed-and-corrected annotations}
\label{table:5}
\end{table}

\subsection{Second classification model}

Combining the tweets labeled unanimously by the crowdworkers with those labeled by the community annotators yielded a training set with 2,665 gold-standard-labeled ``job-related'' tweets and 3,250 ``not job-related'' tweets. We then trained a new classifier using a support vector machine\footnote{http://scikit-learn.org/stable/modules/generated/\\sklearn.svm.SVC.html}. Since the training data are not class-balanced, we grid-searched on a range of class weights and chose the model that optimized F1 score, using 10-fold cross validation. The parameter settings that gave the best results on the held-out data were a linear kernel with the penalty parameter of the error term C = 0.1 and class weight ratior of 1:1 between the classes. Table \ref{table:6} shows the top features.

\begin{table}[h]
\centering 
\begin{tabular}{|c|c|}
\hline
\multirow{3}{*}{\textbf{Positive}} & work, job, manager, jobs, managers \\
 & working, bosses, lovemyjob, shift, worked \\
 & paid, worries, boss, seriously, money \\ \hline
\multirow{3}{*}{\textbf{Negative}} & did, amazing, nut, hard, constr \\
 & phone, doing, since, brdg, play \\
 & its, think, thru, hand, awesome \\ \hline
\end{tabular}
\caption{Top 15 features in positive and negative classes}
\label{table:6}
\end{table}

To evaluate this second model, we used another held-out data set consisting of 5,200 tweets -- 200 with ``job-related'' and 5,000 with ``not job-related'' gold-standard labels. 
This second model obtained 98\% precision and 93\% recall performance for the positive class (``job-related'') after testing the optimal model on this held-out data set.

We then used this model to classify the tweets in our dataset not used for training or evaluation. Almost 200,000
of these tweets were labeled as ``job-related''.
We ranked these job-related tweets by their confidence scores in descending orders as $\tau_{1}$. We then ranked them by their LIWC ``work'' scores similarly as $\tau_{2}$. We used the Kendall rank correlation coefficient \cite{abdi2007kendall} to measure the rank correlation statistically between these two ranking lists. Our result K($\tau_{1}$, $\tau_{2}$) = -0.055 indicates that these measures are mutually independent. We conjecture that there are two reasons for this independence: ``work'' in LIWC lexicon is a broader category than our focus in this study here -- it comprises a set of words related to school activities, like \textit{``scholar, research, highschool, student, quiz''}, which were intended to be excluded in our data filtering stage. The computational process of LIWC score is another potential cause -- it is derived from the numbers of single words matched in ``work'' category divided by the total words in each message -- consequentially it lost all the contextual information compared to the n-gram features used in our classification model.

\subsection{Separating individual users from others}

 To get a sense of the variety of topics discussed, we manually examined the tweets labeled by this process as job-related. A number of tweets are for job openings or personnel recruitment ads, posted by companies or commercial agents, for example ``\emph{Panera Bread: Baker - Night (\#Rochester, NY) http://URL \#Hospitality \#VeteranJob \#Job \#Jobs \#TweetMyJobs}.'' We searched for tweets with similar patterns and then divided the job-related tweets into two subclasses: those from individual users and from commercial users. Basic lexical differences between these two classes are summarized in Table \ref{two_groups}.

\begin{table*}[ht]
\centering
\begin{tabular}{|c|c|c|}
\hline
\textbf{} & \textbf{Individual users group} & \textbf{Commercial users group} \\ \hline
\textbf{Total number of tweets} & 119,376 & 17,641 \\ \hline
\textbf{Total number of unique accounts} & 80,537 & 227 \\ \hline
\textbf{Total number of tokens} & 1,837,304 & 1,400,647 \\ \hline
\textbf{Average number of tokens per tweet} & 15.391 & 17.391 \\ \hline
\textbf{Total number of unique tokens} & 103,089 & 22,547 \\ \hline
\textbf{Average number of unique tokens per tweet} & 0.864 & 0.280 \\ \hline
\textbf{Unique tokens : tokens ratio} & 0.056 & 0.016 \\ \hline
\textbf{Number of hapax legomena} & 69,542 & 7,884 \\ \hline
\textbf{Average number of hapax legomena per tweet} & 0.583 & 0.098 \\ \hline
\end{tabular}
\caption{Basic lexical statistics comparisons between the two groups. \textit{hapax legomena} are those tokens that appear only once in the dataset.}
\label{two_groups}
\end{table*}

The TweetNLP POS tagger (with the Penn Treebank-style tagset) was used to explore different structural attributes between individual users group and commercial users group. The POS tagger assigns parts of speech at a fine-grained level to words used in different contexts accordingly. %

\subsection{Measuring individuals' affective attributes}

We measured seasonal and diurnal variations in mood in job-related discourse. We considered two affective LIWC dimensions: positive affect (PA) and negative affect (NA). These two dimensions are defined as the ratios of the numbers of words in each tweet that are in the PA/NA LIWC lexica to the total number of words in the tweet.

\subsection{Topic model analysis}

Another part of content analysis is modeling the topics hidden in the job-related messages at individual users level. Topic models are a suite of algorithms which enable us summarize and discover thematic information about job-related focuses, interests and trends. We used latent Dirichlet allocation (LDA) \cite{Blei:2012:PTM:2133806.2133826}. The intuitions behind LDA include that a number of ``topics'' are distributed over the words in the whole collection of documents. We aggregated the tweets posted by the same user as a single document. We used the Gensim implementation of LDA \cite{rehurek_lrec} with default settings and 20 topics, with the number of topics chosen empirically, based on experimental results.

\subsection{Time series analysis}

The data we collected from DataSift use the Coordinated Universal Time standard (UTC) to record when each message was created. Timestamps were converted to local time zone 
with daylight saving time taken into consideration. The time series analysis relies on the local time at which each message was posted. 

\section{Results and Discussion}

\subsection{POS tagging comparisons}

\begin{figure*}[ht]
\includegraphics[width=\linewidth]{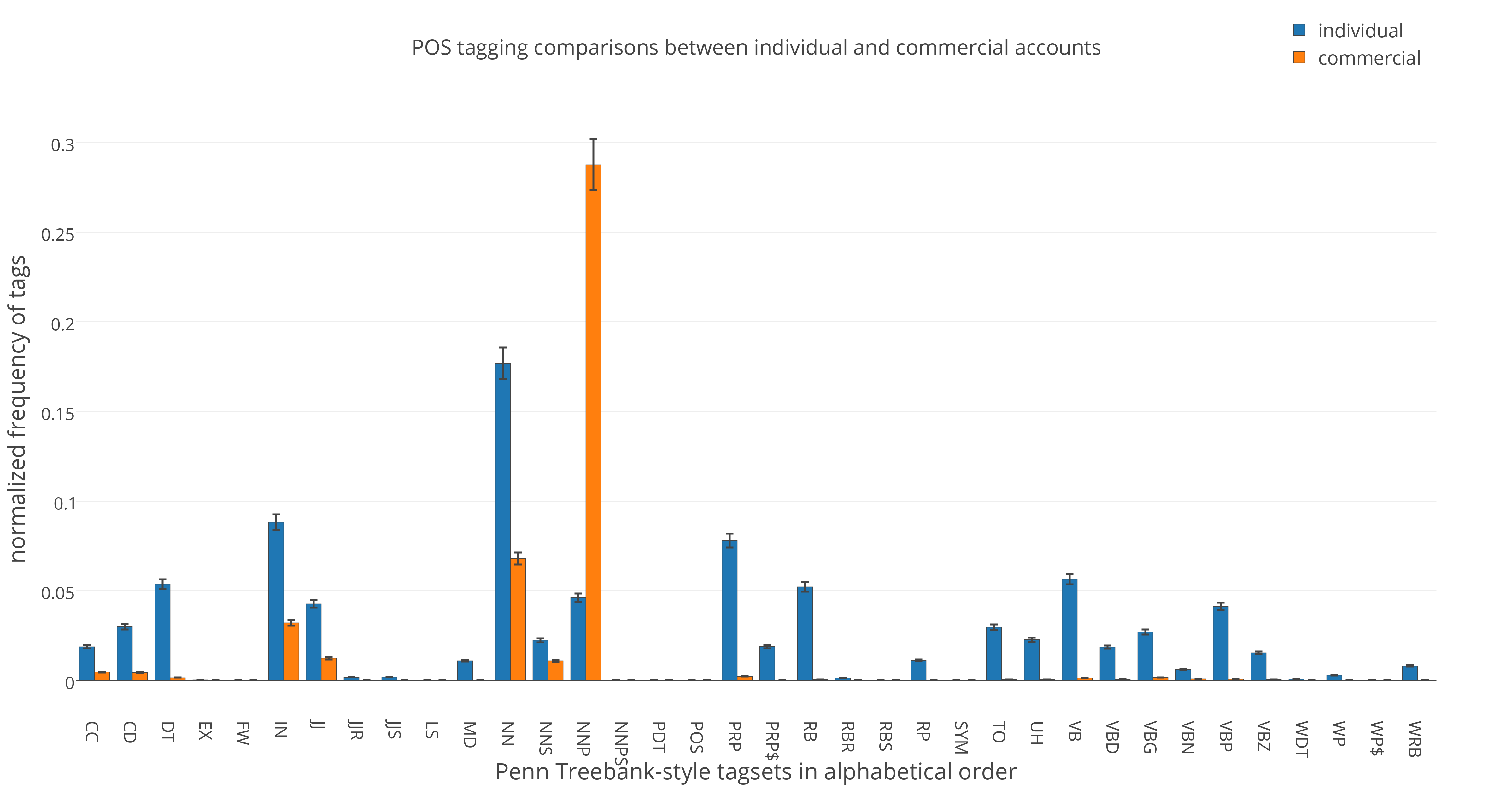}
\caption{POS tagging comparisons between individual and commercial accounts}\label{POS_comparisons}
\centering
\end{figure*}

Figure \ref{POS_comparisons} shows the POS tagging comparisons between individual and commercial users groups. It describe a total of 36 different part-of-speech tags\footnote{The relevant abbreviations were borrowd from \cite{santorini1990part}.}  with average frequencies of each tag for different users groups after being normalized.

The individual users group use CC, CD, DT, IN, JJ, NN, NNS, PRP, PRP\$, RB, RP, TO, UH, VB, VBD, VBG, VBP, VBZ and WRB more frequently than the commercial users group does. The only attribute that the commercial users group surpasses the individual counterpart is NNP. Both groups have barely the rest items detected in each language usage. 

The commercial users group use many NNP, for example \textit{``New York, Accountant, Apple''} in their posts, which supports our assumptions that this group of accounts posted quantities of job openings or advertisements with names or job titles mentioned to give general descriptions. Compared to that, the individual users used the NNP less frequently and in a casual way, like \textit{``Jojo, galactica, Valli''}.

The most frequent tag used by the individual users was NN: e.g. \textit{``application, check, efficiency''}. The second broadly-used tag in the individual users group was IN: for instance \textit{``causee, @, backto, cuz''}. Other tags used heavily by individual users were illustrated as the following tag and samples pairs. CC --- \textit{``aaaaaaand, Buttttt, yeeeeet''}; CD --- \textit{``\$12.50, 9am-10pm, twoooo''}; DT --- \textit{``Whose, thissss, Yahoo's''}; JJ --- \textit{``PROUD, greeeeeat, Hhhhhhaaaaaaappppppyyyyyy''}; NNS --- \textit{``Weekss, bbqs, complainers''}; PRP --- \textit{``imma, yourselves, watcha''}; RB --- \textit{``Sadly, tirelessly, FINALLY''}; TO --- \textit{``To, 2keep, t0''}; UH --- \textit{``Yayyyyy, Ahahaha, lololol''}; VB --- \textit{``git, guarantee, re-do''}; VBD --- \textit{``Upset, planned, debated''}; VBG --- \textit{``tryin, working, starvin''}; VBP --- \textit{``hate, harass, do''}; WRB --- \textit{``YYYYYYY, Wot, where''}

\subsection{Topic analysis}

We performed LDA topic analysis to determine what individual users particularly talked about in job-related messages. %
We observed that several topics show notable signals about job-related theme. See Table \ref{topics_words}.

\begin{table}[ht]
\centering
\begin{tabular}{|c|c|}
\hline
\textbf{topic number} & \textbf{representative words} \\ \hline
\textbf{Topic 4} & \begin{tabular}[c]{@{}c@{}}tomorrow, working, today, week, \\ monday, time, weekend, day, \\ minute, hour, morning, night\end{tabular} \\ \hline
\textbf{Topic 6} & \begin{tabular}[c]{@{}c@{}}accept, canceled, trust, quit, \\ working, support, \#struggling,\\ unemployed, helping, corporate, \\ planning, professional\end{tabular} \\ \hline
\textbf{Topic 14} & \begin{tabular}[c]{@{}c@{}}ugh, exhausted, feeling,\\ competition, effort, celebrate\end{tabular} \\ \hline
\textbf{Topic 20} & \begin{tabular}[c]{@{}c@{}}technician, \#jobs, manager, \\ productivity, contractor, associate, \\ assistant, intern, industry\end{tabular} \\ \hline
\end{tabular}
\caption{Example of topics and corresponding representative words}
\label{topics_words}
\end{table}
Among the more salient topics, 
topic 4 contains notions of time related to routine work. Topic 6 mixes the rise and fall statuses about career life. Topic 14 manifests the conceivable tensions and challenges at work. Topic 20 illustrates diverse occupations and roles in work force.

\subsection{Analysis of Twitter usage}

\begin{figure}[h]
\includegraphics[width=\linewidth]{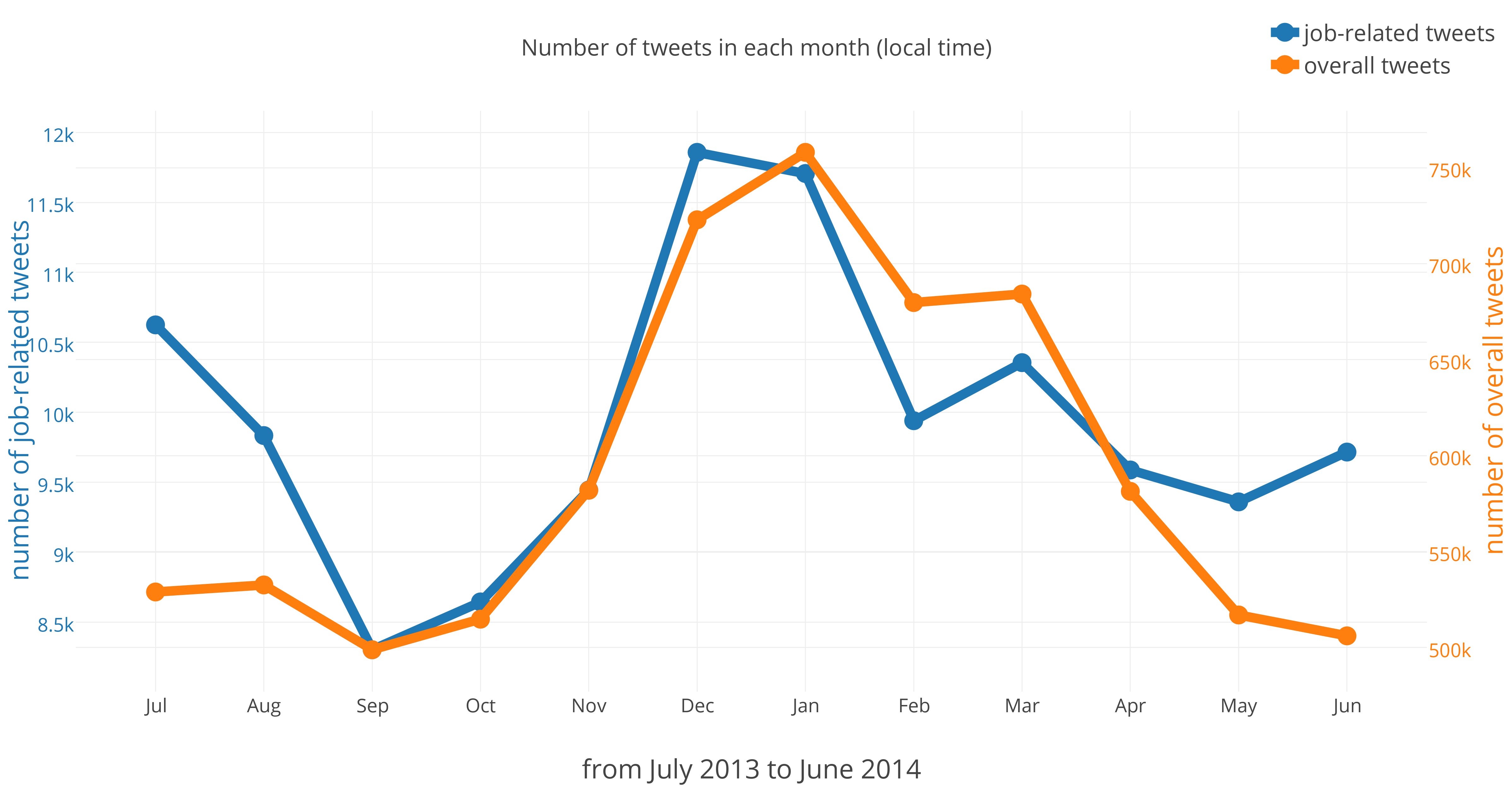}
\caption{Numbers of tweets in each month}\label{month_count}
\centering
\end{figure}

Figure \ref{month_count} shows that the total number of tweets per month and the number of job-related tweets per month follow similar seasonal trends, though the overall count peaks in January and the job-related count peaks in December. The overall count and the job-related count both drop to the bottom in September.

\begin{figure}[h!]
\includegraphics[width=\linewidth]{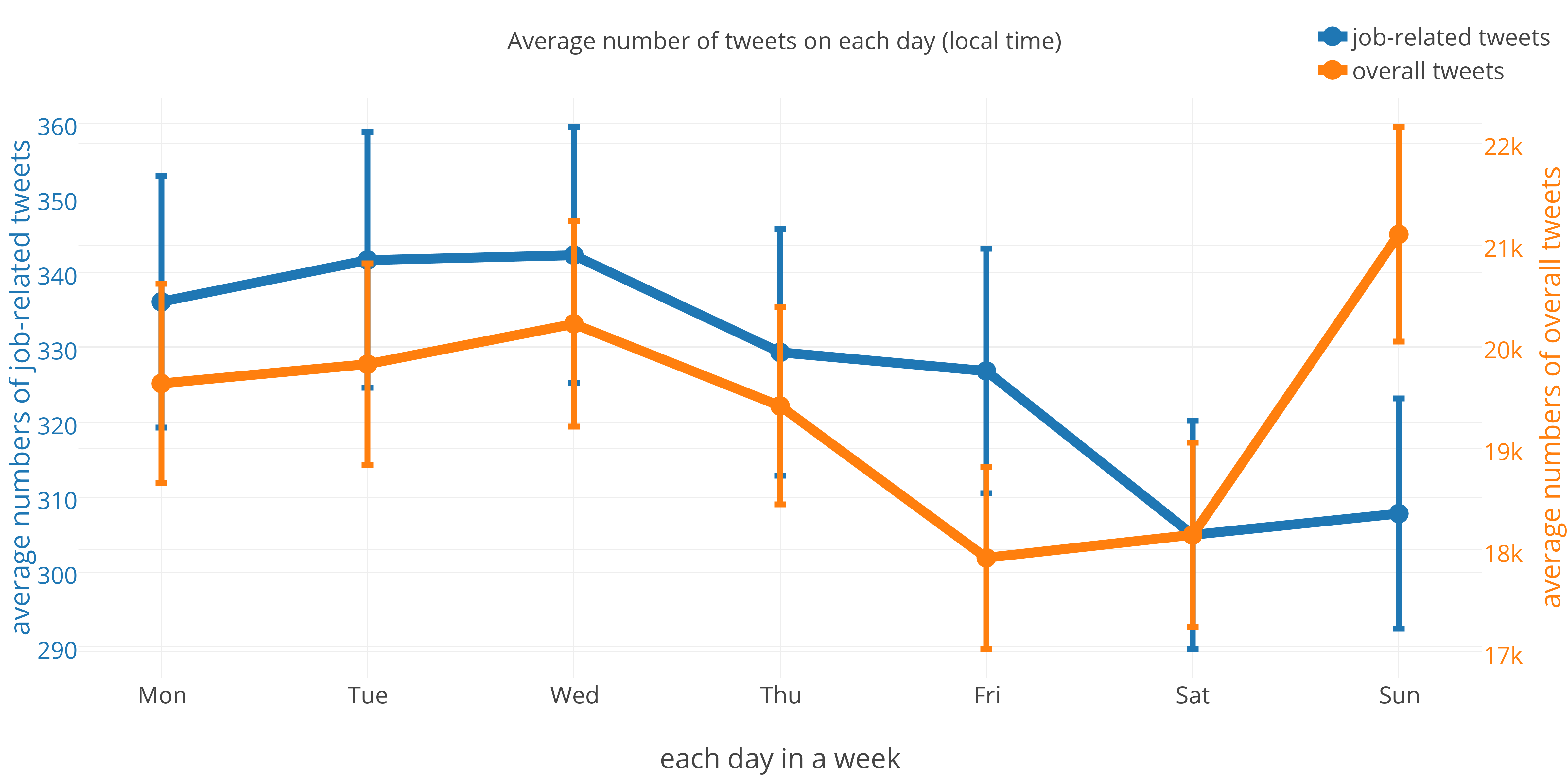}
\caption{Average numbers of tweets on each day of week}\label{week_count}
\centering
\end{figure}

Figure \ref{week_count} shows weekly trends of both overall tweets and the job-related tweets. The average number of job-related tweets starts steadily on Monday, peaks on Wednesday and decreases gradually until bottoming out on Saturday, and stays stable until Sunday, which follows the standard work week periodicity. Sunday had the largest volume of tweets -- greatly exceeding the job-related tweets -- many of which were related to active social activities. Friday and Saturday were the least active days from online interactions perspective. 

\begin{figure}[ht]
\includegraphics[width=\linewidth]{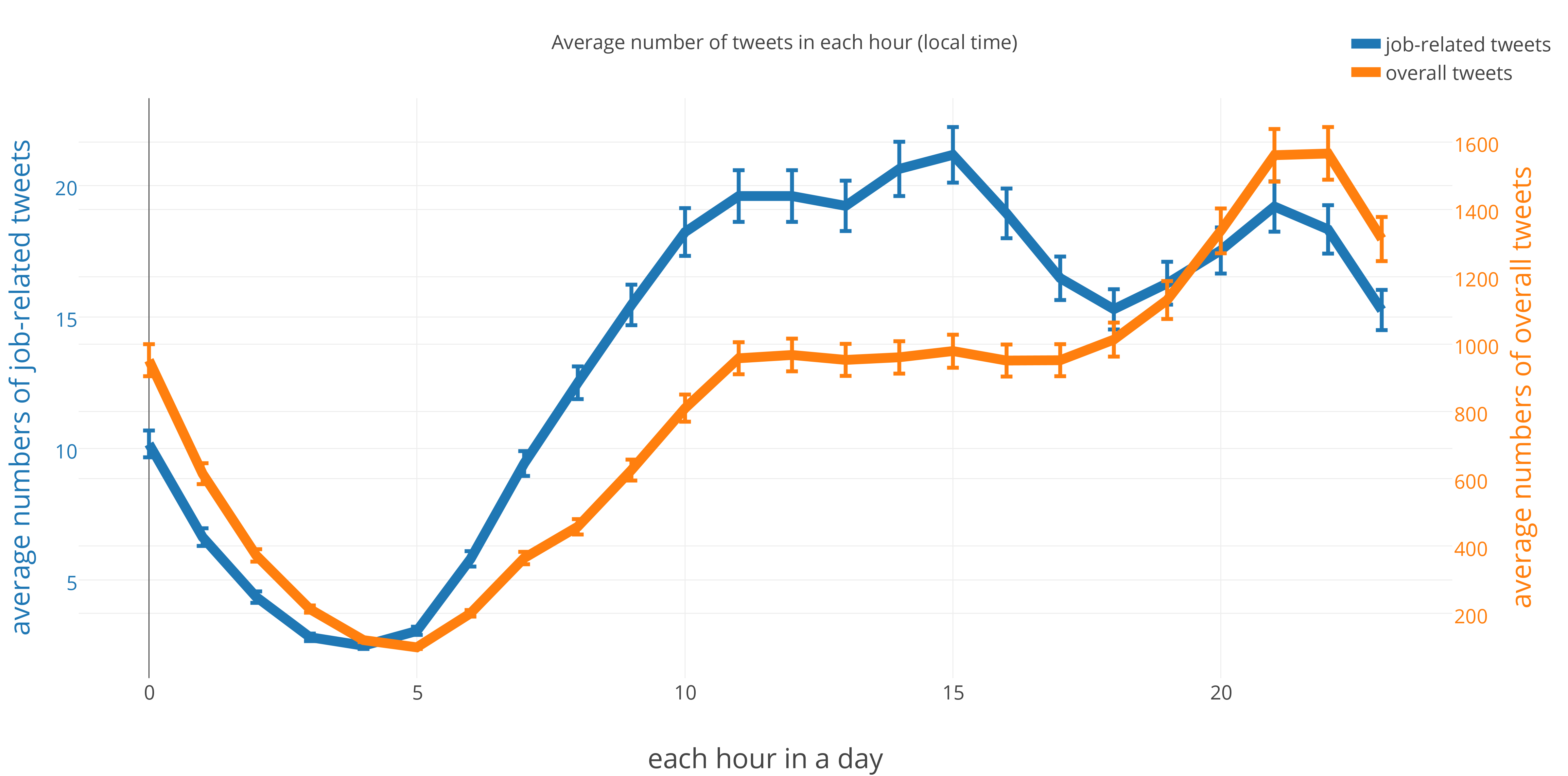}
\caption{Average numbers of tweets in each hour}\label{hour_count}
\centering
\end{figure}

Figure \ref{hour_count} shows daily trends in volume. Job and overall trends ran parallel before 5 o'clock and then diverged. The average number of job-related tweets increased faster than the volume of overall tweets, until 3pm. This suggests that people posted more job-related tweets in morning and early afternoon. The average number of job-related tweets sharply decreased until 6pm. After that, it increases modestly and reaches another high point around 9pm. The average overall volume of tweets peaked at 9 and 10 in the evening.

\subsection{Affective changes observations}

To observe the affective changes and temporal correlations hidden behind job-related tweets, we accumulated the positive affect (PA) and negative affect (NA) for the job-related tweets in each hour on different days in a week.

\begin{figure}[h]
\includegraphics[width=\linewidth]{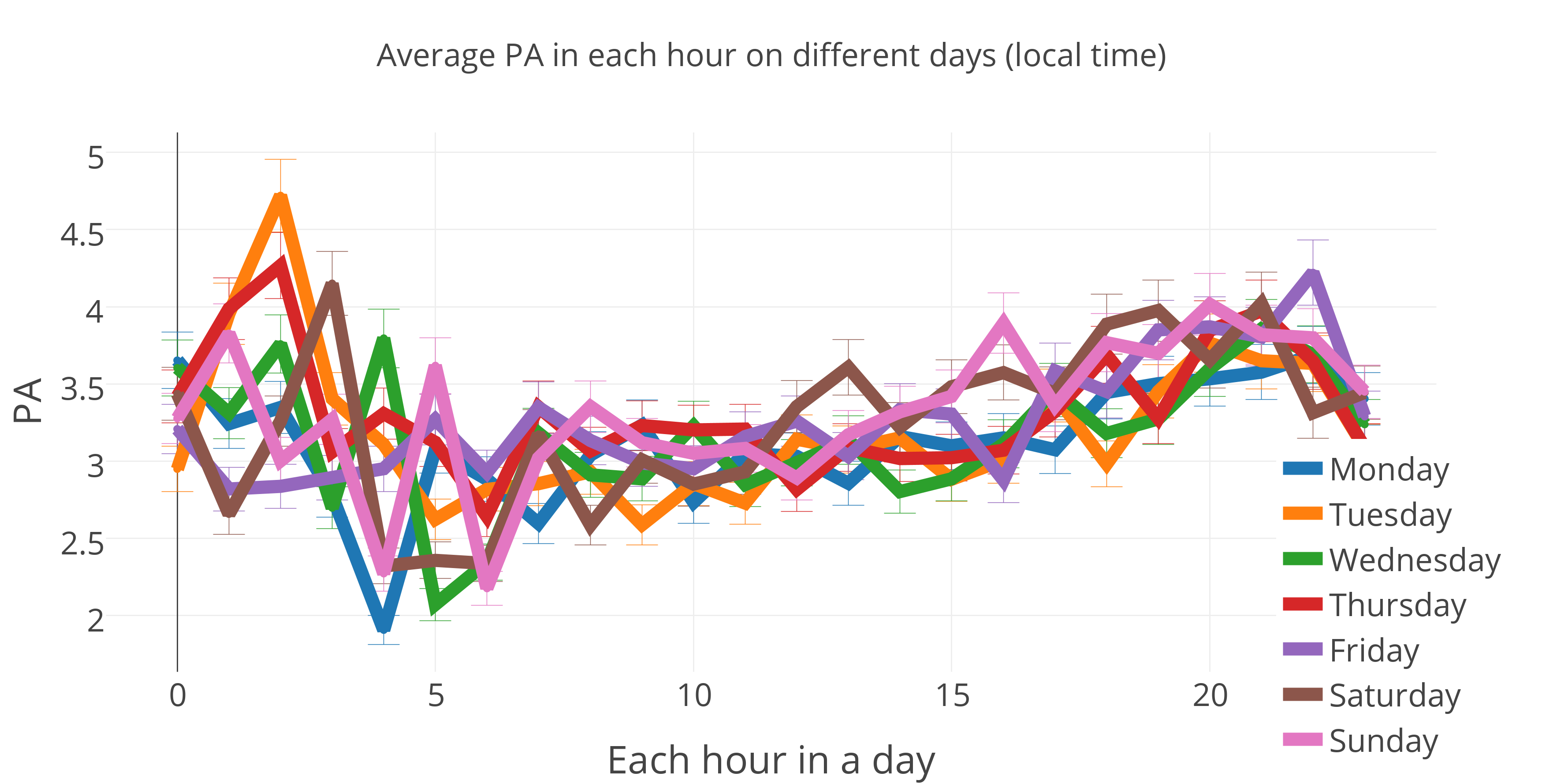}
\caption{Hourly PA changes broken down by different days of the week}\label{avg_PA}
\centering
\end{figure}

\begin{figure}[h]
\includegraphics[width=\linewidth]{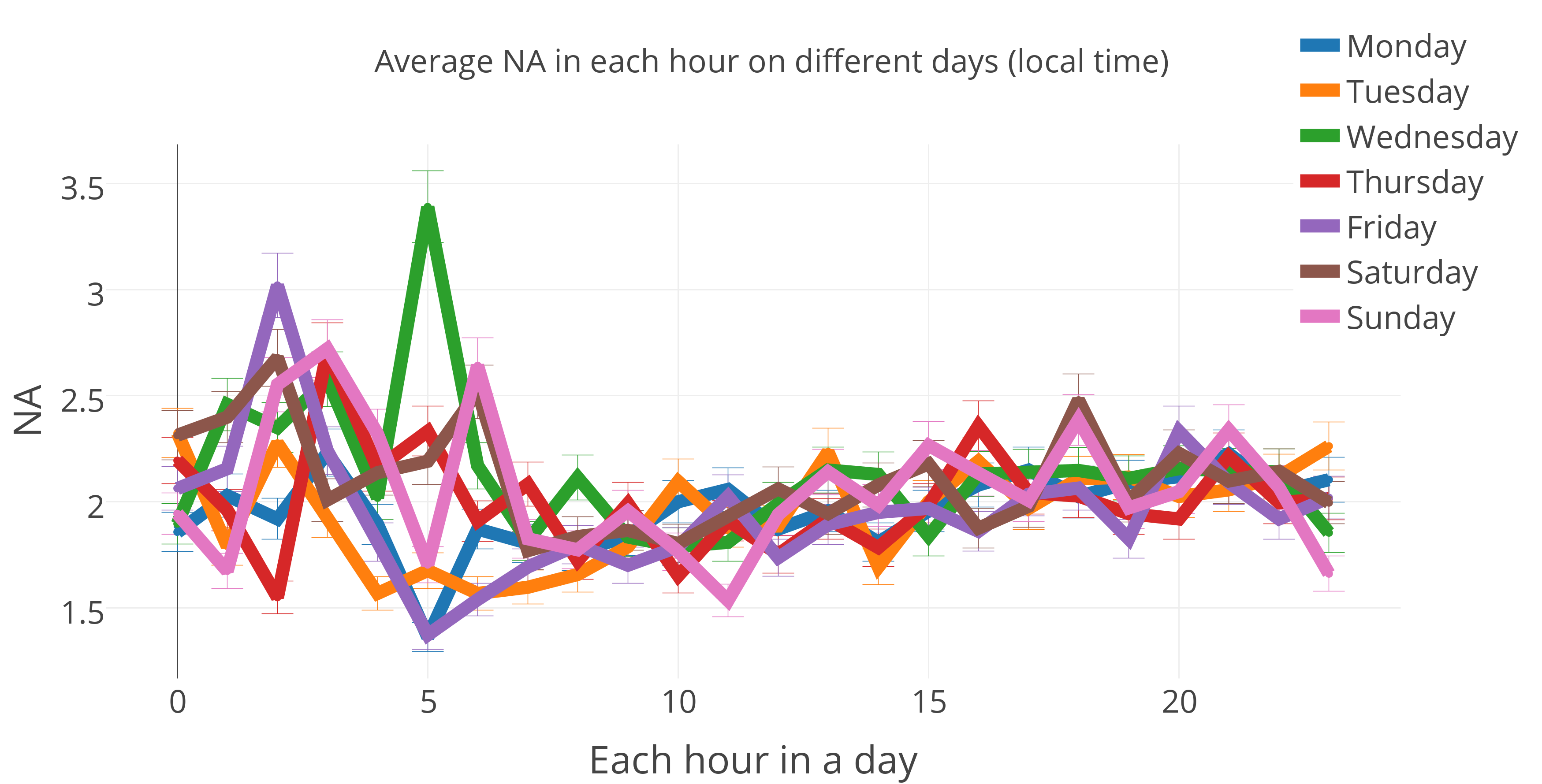}
\caption{Hourly NA changes broken down by different days of the week}\label{avg_NA}
\centering
\end{figure}

Figure \ref{avg_PA} and \ref{avg_NA} show hourly and daily changes of average PA and NA for individual users group in local time, with 95\% confidence intervals. Both PA and NA affect changes have fluctuations during each day and share the similar shape respectively across days of the week.

PA levels are generally higher on weekends (Saturday and Sunday) than other days during the week. Tuesdays and Thursdays witness the PA peak at 2am in the morning. And PA bottoms out at 4am on Mondays and 5am on Wednesdays. 

NA has its highest point at 5am on Wednesdays, then at 2am on Fridays. Relatively to PA, NA does not change inversely which indicates that PA and NA vary independently and are not mutually exclusive. 

\section{Conclusion}

We used crowdsourcing techniques and local expertise to power a supervised learning pipeline that iteratively improves the classification accuracy of job-related tweets. Using this fine-grained text-based classification model, we extracted high quality job-related tweets from our local region. We separated commercial accounts from individual accounts and measured psychological states for individual users using LIWC.

Our findings show that even though jobs take up an enormous amount of most adults' time, job-related tweets are rather infrequent --- about 1\% to 2\% of overall tweets (see Figure \ref{month_count}). Inspecting the usage patterns of Twitter on each day of the week, we find that Sunday is busiest and Friday the quietest. People post most job-related messages on Wednesday, and tweet much less about jobs on the weekends. The volume of job-related tweets starts increasing from 5am each day and reaches the peak at 3pm. It is interesting to see another increase of job-related tweets after 6pm until 9pm.

We examined affective changes in job-related tweets --- primarily PA and NA --- in daily and hourly settings and concluded that PA and NA change independently, thus, e.g., low NA indicates the absence of negative feelings, not the presence of positive feelings. Usually tweets on weekends convey higher PA and NA than those on weekdays.

Our work has several limitations. The data are not massive enough to conduct year-to-year comparison studies on seasonal job-related trends. This work is a preliminary exploration that relies heavily on linguistic models built upon the manual annotations. We have not examined whether providing contextual information in annotation tasks would influence the model performance. Also due to the demographic characteristics of Twitter, we are less likely to observe working senior citizens. Future research would benefit from more tightly integrated quantitative and qualitative analyses, such as geographical analysis of the job-related data in local communities.

\section{Acknowledgments}
This work was supported in part by a GCCIS Kodak Endowed Chair Fund Health Information Technology Strategic Initiative Grant and NSF Award \#SES-1111016.

\balancecolumns


\begin{thebibliography}{10}

\bibitem{abdi2007kendall}
H.~Abdi.
\newblock {T}he {K}endall {R}ank {C}orrelation {C}oefficient.
\newblock {\em Encyclopedia of Measurement and Statistics. Sage, Thousand Oaks,
  CA}, pages 508--510, 2007.

\bibitem{Blei:2012:PTM:2133806.2133826}
D.~M. Blei.
\newblock {P}robabilistic {T}opic {M}odels.
\newblock {\em Commun. ACM}, 55(4):77--84, Apr. 2012.

\bibitem{bollen2011twitter}
J.~Bollen, H.~Mao, and X.~Zeng.
\newblock {T}witter {M}ood {P}redicts {T}he {S}tock {M}arket.
\newblock {\em Journal of Computational Science}, 2(1):1--8, 2011.

\bibitem{timeuse}
{B}ureau of~{L}abor {S}tatistics.
\newblock Time use on an average work day for employed persons ages 25 to 54
  with children, 2013.

\bibitem{callison2009fast}
C.~Callison-Burch.
\newblock {F}ast, {C}heap, and {C}reative: {E}valuating {T}ranslation {Q}uality
  {U}sing {A}mazon's {M}echanical {T}urk.
\newblock In {\em Proceedings of the 2009 Conference on Empirical Methods in
  Natural Language Processing: Volume 1-Volume 1}, pages 286--295. Association
  for Computational Linguistics, 2009.

\bibitem{suicidefigure}
{Centers for Disease Control and Prevention}.
\newblock {C}ost {E}stimates of {V}iolent {D}eaths: {F}igures and {T}ables,
  2013.

\bibitem{cheng2010you}
Z.~Cheng, J.~Caverlee, and K.~Lee.
\newblock {Y}ou {A}re {W}here {Y}ou {T}weet: {A} {C}ontent-{B}ased {A}pproach
  to {G}eo-{L}ocating {T}witter {U}sers.
\newblock In {\em Proceedings of the 19th ACM international conference on
  Information and knowledge management}, pages 759--768. ACM, 2010.

\bibitem{socialtop}
comScore.
\newblock {I}t's a {S}ocial {W}orld: {T}op 10 {N}eed-to-{K}nows {A}bout
  {S}ocial {N}etworking and {W}here {I}t's {H}eaded, 2011.

\bibitem{de2013understanding}
M.~De~Choudhury and S.~Counts.
\newblock {U}nderstanding {A}ffect in the {W}orkplace via {S}ocial {M}edia.
\newblock In {\em Proceedings of the 2013 conference on Computer supported
  cooperative work}, pages 303--316. ACM, 2013.

\bibitem{de2013predicting}
M.~De~Choudhury, M.~Gamon, S.~Counts, and E.~Horvitz.
\newblock {P}redicting {D}epression via {S}ocial {M}edia.
\newblock In {\em AAAI Conference on Weblogs and Social Media}, 2013.

\bibitem{evanini2010using}
K.~Evanini, D.~Higgins, and K.~Zechner.
\newblock {U}sing {A}mazon {M}echanical {T}urk for {T}ranscription of
  {N}on-{N}ative {S}peech.
\newblock In {\em Proceedings of the NAACL HLT 2010 Workshop on Creating Speech
  and Language Data with Amazon's Mechanical Turk}, pages 53--56. Association
  for Computational Linguistics, 2010.

\bibitem{fleiss1971measuring}
J.~L. Fleiss.
\newblock {M}easuring {N}ominal {S}cale {A}greement {A}mong {M}any {R}aters.
\newblock {\em Psychological Bulletin}, 76(5):378, 1971.

\bibitem{gallup}
Gallup.
\newblock {S}tate of the {A}merican {W}orkplace, 2013.

\bibitem{golder2011diurnal}
S.~A. Golder and M.~W. Macy.
\newblock {D}iurnal and {S}easonal {M}ood {V}ary with {W}ork, {S}leep, and
  {D}aylength {A}cross {D}iverse {C}ultures.
\newblock {\em Science}, 333(6051):1878--1881, 2011.

\bibitem{worksuicide}
{{H}azards {M}agazine}.
\newblock {W}ork {S}uicide, 2014.

\bibitem{iKeepSafe}
iKeepSafe.
\newblock {S}uicide: {U}sing {T}echnology for {D}etection and {I}ntervention,
  2014.
\newblock [Online; accessed 9-December-2014].

\bibitem{klaus1980content}
K.~Klaus.
\newblock {C}ontent {A}nalysis: {A}n {I}ntroduction to {I}ts {M}ethodology,
  1980.

\bibitem{tong2015detecting}
T.~Liu, C.~M.~Homan, C.~Ovesdotter~Alm, A.~Marie~White, M.~C.~Lytle-Flint, and
  H.~A.~Hkutz.
\newblock {D}etecting {J}ob-related {M}essages in {T}witter.
\newblock Submitted.

\bibitem{nasukawa2003sentiment}
T.~Nasukawa and J.~Yi.
\newblock {S}entiment {A}nalysis: {C}apturing {F}avorability {U}sing {N}atural
  {L}anguage {P}rocessing.
\newblock In {\em Proceedings of the 2nd international conference on Knowledge
  capture}, pages 70--77. Association for Computing Machinery, 2003.

\bibitem{oxman1982language}
T.~E. Oxman, S.~D. Rosenberg, and G.~J. Tucker.
\newblock {T}he {L}anguage of {P}aranoia.
\newblock {\em The American Journal of Psychiatry}, 1982.

\bibitem{pak2010twitter}
A.~Pak and P.~Paroubek.
\newblock {T}witter as a {C}orpus for {S}entiment {A}nalysis and {O}pinion
  {M}ining.
\newblock In {\em Language Resources and Evaluation}, 2010.

\bibitem{paul2011you}
M.~J. Paul and M.~Dredze.
\newblock {Y}ou {A}re {W}hat {Y}ou {T}weet: {A}nalyzing {T}witter for {P}ublic
  {H}ealth.
\newblock In {\em International Conference on Weblogs and Social Media}, 2011.

\bibitem{pirzadeh2012emotion}
A.~Pirzadeh and M.~S. Pfaff.
\newblock {E}motion {E}xpression under {S}tress in {I}nstant {M}essaging.
\newblock In {\em Proceedings of the Human Factors and Ergonomics Society
  Annual Meeting}, volume~56, pages 493--497. Sage Publications, 2012.

\bibitem{poulin2014predicting}
C.~Poulin, B.~Shiner, P.~Thompson, L.~Vepstas, Y.~Young-Xu, B.~Goertzel,
  B.~Watts, L.~Flashman, and T.~McAllister.
\newblock {P}redicting the {R}isk of {S}uicide by {A}nalyzing the {T}ext of
  {C}linical {N}otes.
\newblock {\em PLOS ONE}, 9(1):e85733, 2014.

\bibitem{rao2010classifying}
D.~Rao, D.~Yarowsky, A.~Shreevats, and M.~Gupta.
\newblock {C}lassifying {L}atent {U}ser {A}ttributes in {T}witter.
\newblock In {\em Proceedings of the 2nd international workshop on Search and
  mining user-generated contents}, pages 37--44. ACM, 2010.

\bibitem{rehurek_lrec}
R.~{\v R}eh{\r u}{\v r}ek and P.~Sojka.
\newblock {S}oftware {F}ramework for {T}opic {M}odelling with {L}arge
  {C}orpora.
\newblock In {\em {Proceedings of the LREC 2010 Workshop on New Challenges for
  NLP Frameworks}}, pages 45--50, Valletta, Malta, May 2010. ELRA.

\bibitem{rude2004language}
S.~Rude, E.-M. Gortner, and J.~Pennebaker.
\newblock {L}anguage {U}se of {D}epressed and {D}epression-{V}ulnerable
  {C}ollege {S}tudents.
\newblock {\em Cognition \& Emotion}, 18(8):1121--1133, 2004.

\bibitem{sadilek2013modeling}
A.~Sadilek, C.~Homan, W.~S. Lasecki, V.~Silenzio, and H.~Kautz.
\newblock {M}odeling {F}ine-{G}rained {D}ynamics of {M}ood at {S}cale.
\newblock In {\em Workshop on Diffusion Networks and Cascade Analytics in Web
  Search and Data Mining}, 2014.

\bibitem{sadilek2012modeling}
A.~Sadilek, H.~A. Kautz, and V.~Silenzio.
\newblock {M}odeling {S}pread of {D}isease from {S}ocial {I}nteractions.
\newblock In {\em International Conference on Weblogs and Social Media}, 2012.

\bibitem{santorini1990part}
B.~Santorini.
\newblock {P}art-of-{S}peech {T}agging {G}uidelines for the {P}enn {T}reebank
  {P}roject (3rd {R}evision).
\newblock 1990.

\bibitem{schrading2015whyistayed}
N.~Schrading, C.~Ovesdotter~Alm, R.~Ptucha, and C.~Homan.
\newblock \#{W}hyistayed, \#{W}hyileft: {M}icroblogging to {M}ake {S}ense of
  {D}omestic {A}buse.
\newblock In {\em Human Language Technologies: The 2015 Annual Conference of
  the North American Chapter of the ACL}, pages 1281--1286, 2015.

\bibitem{snow2008cheap}
R.~Snow, B.~O'Connor, D.~Jurafsky, and A.~Y. Ng.
\newblock {C}heap and {F}ast -- {B}ut is it {G}ood? {E}valuating {N}on-{E}xpert
  {A}nnotations for {N}atural {L}anguage {T}asks.
\newblock In {\em Proceedings of the conference on empirical methods in natural
  language processing}, pages 254--263. Association for Computational
  Linguistics, 2008.

\bibitem{tamersoy2015characterizing}
A.~Tamersoy, M.~De~Choudhury, and D.~H. Chau.
\newblock {C}haracterizing {S}moking and {D}rinking {A}bstinence from {S}ocial
  {M}edia.
\newblock In {\em Proceedings of the 26th ACM Conference on Hypertext \& Social
  Media}, pages 139--148. ACM, 2015.

\bibitem{tausczik2010psychological}
Y.~R. Tausczik and J.~W. Pennebaker.
\newblock {T}he {P}sychological {M}eaning of {W}ords: {LIWC} and {C}omputerized
  {T}ext {A}nalysis {M}ethods.
\newblock {\em Journal of Language and Social Psychology}, 29(1):24--54, 2010.

\bibitem{tumasjan2010predicting}
A.~Tumasjan, T.~O. Sprenger, P.~G. Sandner, and I.~M. Welpe.
\newblock {P}redicting {E}lections with {T}witter: {W}hat 140 {C}haracters
  {R}eveal about {P}olitical {S}entiment.
\newblock {\em ICWSM}, 10:178--185, 2010.

\end{thebibliography}
\end{document}